\documentclass{iaus}
\usepackage{graphicx}
\def\Msun{M$_\odot$}

\title[The Horizontal Branch and Multiple stellar populations in GCs] 
{Multiple stellar populations in Globular Clusters: collection of information
from the Horizontal Branch}

\author[F.~D'Antona \& V.~Caloi]   
{Francesca D'Antona$^1$~~%
\and ~~Vittoria Caloi$^2$}

\affiliation{$^1$INAF - Osservatorio Astronomico  di Roma, via 
             Frascati 33, I-00040 Monte Porzio, Italy
\break email: dantona@oa-roma.inaf.it \\[\affilskip]
$^2$INAF - IASF, via Fosso del Cavaliere, 
             I-00133 Roma, Italy \break email: vittoria.caloi@iasf-roma.inaf.it}

\pubyear{2007}
\volume{246}  
\pagerange{119--126}
\date{?? and in revised form ??}
\jname{Proceedings of IAU Symposium ``Dynamical Evolution of Dense Stellar Systems"}
\editors{E. Vesperini, M. Gierzs \& A. Sills, eds.}
\begin{document}

\maketitle

\begin{abstract}
The majority of the inhomogeneities in the chemical composition of 
Globular Cluster (GC) stars appear due 
to primordial enrichment by hot-CNO cycled material processed in
stars belonging to a first stellar generation. Either massive AGB envelopes
subject to hot bottom burning, or the envelopes of massive fastly rotating stars
could be the progenitors. In both cases, the
stars showing chemical anomalies must have also enhanced helium abundance, 
and we have proposed that this higher helium
could be at the basis of the many different
morphologies of GC horizontal branches (HB) for similar ages and metallicities.
The helium variations have been beautifully confirmed by the splitting of the main sequence in 
the clusters $\omega$ Cen and NGC 2808, but this effect can show up 
only for somewhat extreme helium abundances. Therefore it is
important to go on using the HB morphology
to infer the number ratio of the first to the second generation in as many clusters 
as possible. We exemplify how it is possible to infer the presence of a He--rich 
stellar component in different clusters thanks to different HB features (gaps, RR Lyr
periods and period distribution, ratio of blue to red stars, blue tails). 
In many clusters at least 50\% of the stars belong
to the second stellar generation, and in some cases we suspect that the stars might all
belong to the second generation.
We shortly examine the problem of the initial mass function required to achieve the observed
number ratios and conclude that: 1) the initial cluster must have been much more massive
than today's cluster, and 2) formation of the second stellar generation mainly in the
central regions of the cluster may help in obtaining the desired values.
\end{abstract}

\firstsection 
\section{Introduction}

The observations of GC stars are still to be interpreted in a fully consistent frame. 
Nevertheless, a general consensus is emerging on the fact that most GCs can not
be considered any longer ``simple stellar populations", and that
``self--enrichment" is a common feature among GCs. This consensus has been the
consequence of three independent lines of evidence:
\begin{itemize}
\item Spectroscopic observations: the discovery of ``chemical anomalies", such as the
Na--O and Mg--Al anticorrelation, dates back to
the seventies. The anomalies are now observed also at the turnoff (TO) 
and among the subgiants (e.g.,\cite[Gratton et al. 2001]{gratton2001},
\cite[Briley et al. 2002, 2004]{briley2002, briley2004}), so they must be
attributed to some process of ``self--enrichment" occurring at the first
stages of the cluster life. There must have been a first epoch 
of star formation that gave origin to the ``normal" (first generation) stars,
with CNO and other abundances similar to the population II field stars of the
same metallicity. Afterwards, there must have been
some other epoch of star formation, including material 
heavily processed through the CNO cycle.
This material either was entirely ejected by stars belonging to the 
first stellar generation, or it is a mixture of ejected and pristine
matter of the initial star forming cloud. We can derive this conclusion as a 
consequence of the fact that 
there is no appreciable difference in the metallicity of the ``normal" and
chemically anomalous stars belonging to the same 
GC. (Needless to say, this statement {\it does not}
hold for $\omega$ Cen, which must indeed be considered a small galaxy and not a typical
GC. In the following, we will only examine ``normal clusters", those which
do not show signs of metal enrichment due to supernovae ejecta). 
This is an important fact that tells us, e.g., that it is highly improbable that 
the chemical anomalies are due to mixing of stars born in two 
different clouds, as there is no reason why the two clouds should 
have a unique metallicity. In addition, the clusters showing
chemical anomalies have a huge variety of metallicities, making the suggestion
of mixing of two different clouds even more improbable.
The matter must have been processed through the hot CNO cycle, and not, or only marginally,
through the helium burning phases, as the sum of CNO elements is the same in the ``normal"
and anomalous stars (\cite[Cohen \& Mel\'endez 2005]{cohenmelendez2005}). The progenitors then may be either massive Asymptotic Giant Branch
(AGB) star (\cite[Ventura et al. 2001, 2002]{ventura2001, ventura2002}) 
or fast rotating massive stars (\cite[Decressin et al. 2007]{decressin}).

\item Interpretation of the Horizontal Branch morphology in terms of helium 
content variations among the GC stars.
Whichever the progenitors, some helium enrichment must be present in the matter processed
through hot CNO. \cite{dantona2002} recognized that this could have a strong effect on the
Horizontal Branch (HB) morphology, and even help to explain some features (gaps, hot blue
tails, second parameter effect) which had defied all reasonable alternative explanation. 
A wide variety of problems has been examined in the latest years: the very peculiar 
morphology of the HB in the massive cluster 
NGC~2808, interpreted in terms of varying helium among its stars; the
second parameter effect in the clusters M~13 and M~3 (\cite[Caloi \& D'Antona, 2005]
{caloi-dantona2005}); the very peculiar features of the GCs NGC~6441 and NGC~6388
(\cite[Caloi \& D'Antona, 2007a]{caloi-dantona2007a}), which can all be modeled by
assuming that quite a large fraction of stars have high helium.
Hints that this was correct came from the analysis of the helium content in hot
HB stars (\cite[Moehler et al. 2004]{moeler2004}).

\item Photometric splitting of the main sequence in a few clusters. The 
necessity of a varying helium content in NGC~2808 was later confirmed by the first analysis 
of the cluster main sequence (\cite[D'Antona et al. 2005]{dantona2005}), that showed
a tail of ``blue" stars. This
could only be interpreted as a very helium rich sequence. In fact, 
\cite{carretta2006} had shown that the metallicity of oxygen poor and oxygen normal stars
is the same. The recent new HST observations by \cite{piotto2007} leave no doubt that there
are at least three different populations in this cluster. This came after the first 
discovery of a peculiar blue main sequence in $\omega$~Cen (\cite[Bedin et al. 2004]
{bedin2004}), interpreted 
again in terms of a very high helium content (\cite[Norris 2004]{norris2004}, \cite[Piotto
et al. 2005]{piotto2005}).
The HB observations however have shown that the blue main sequences are only the tip of 
the iceberg of the self--enrichment. In most clusters the higher helium abundances
remain confined below Y$\sim 0.30$, and the presence of such stars will not be clearcut 
from main sequence observations (\cite[D'Antona et al. 2002]{dantona2002}, \cite[Salaris
et al. 2006]{salaris2006}). Viceversa, if we wish to shed light on the entire process of
formation of GCs we must have a rough idea of the total number of chemically
anomalous stars.
\end{itemize}
Consequently, we decided to continue our investigation of the HB in as many
clusters as possible, in the hypothesis that the HB  morphology can be mainly interpreted
in terms of a helium content distribution among the stars. We summarize here some of the
results, as a basis to discuss the initial mass function of the first generation, required 
to produce the needed fraction of stars in the second generation.

\section{Different clusters, different necessity for helium rich stars}\label{sec:2}

We recall that the role of helium rich stars is different according to which is the
basic HB morphology of the first generation stars (\cite[D'Antona et al. 2002]{dantona2002}). 

\subsection{NGC 2808}
If the age, metallicity and 
mass loss are such that normal--helium stars populate a red clump, the stars with
helium enhancement (which are less massive) will populate the bluer HB and the RR Lyr
region. If there is a gap between the normal--helium stars and the {\it minimum} helium 
content of the second generation (a situation which does probably occur, if the 
processed matter comes from the massive AGBs), the case of NGC~2808 shows up: a red
clump (first generation), almost no RR Lyr (due to the helium gap) and a blue part with
larger helium content (Y$\sim 0.28$\ according to \cite[D'Antona and
Caloi 2004]{dantona-caloi2004}).
In addition, the presence of a separate very high helium population -as derived from the main
sequence- may explain the two blue tails of the HB (\cite[D'Antona et al. 2005]{dantona2005}).
The cluster seems to be divided into 50\% normal--helium stars, and 50\% helium enriched
stars, but remember that the very high helium (Y$\sim$0.40) stars are only $\sim$15\%.

\subsection{NGC 6441 and NGC 6388. And also 47~Tuc}
The case of these two high metallicity clusters is even more interesting: here the red
clump extends for about a magnitude thickness, and any attempt to attribute this to
differential reddening has failed (\cite[Raimondo et al. 2002]{raimondo2002}). 
The RR Lyr have a very long
period, unexplicable for the metallicity (\cite[Pritzl et al. 2000]{pritzl2000}). 
And the HB extends even to the region of blue, 
hot stars. \cite{caloi-dantona2007a} show that this is another case study: the morphology
requires not only some helium enrichment for the bluer side of the HB, as we could 
naively think, but extreme helium enrichment {\it even for the red clump stars!} In fact,
the high helium -- high metallicity helium core burning low mass stars make long loops
from red to blue in the HB (\cite[Sweigart \& Gross 1976]{sweigart-gross1976}). 
This is due to the fact that the
higher mean molecular weight --leading to a high H--burning shell temperature-- and the high
metallicity --leading to a stronger CNO shell-- both conspire towards the result 
that the H--shell energy source prevails with respect to the He--core burning. 
The consequent growth of the helium core leads evolution towards the blue. 
Therefore, if we must explain the luminous 
(long period) RR Lyr by stars having high helium, the same stars will also populate the
red clump: this is exactly what we observe: if the helium content is not as large as Y$\sim 0.35$\ 
{\it in the red clump}, the HB finds no satisfactory explanation. The percentage of
helium enriched stars is in this case $\sim$60\% for NGC 6441 and the same for NGC~6388.
The main difference among the two clusters is that NGC~6388 seems to have a higher tail
of very high helium (Y$>$0.35) stars, reaching $\sim$20\%. If we analyze
47~Tuc, the prototype of metallic GCs, the red clump seems to imply higher helium only
for $\sim$25\% of the stars. Why a cluster almost as massive as the other two is so
different --more normal-- remains to be explained.

\begin{table}[ht]
\caption{Helium history of 8 clusters}
\smallskip
\begin{center}
{\small
\begin{tabular}{cc|cc|cc|cc|cc|cc|cc|cc}
\hline
\noalign{\smallskip}
\multicolumn{2}{c}{NGC 2808}& \multicolumn{2}{c}{NGC 6441}& \multicolumn{2}{c}{NGC 6388}
& \multicolumn{2}{c}{47 Tuc}& \multicolumn{2}{c}{M3} & \multicolumn{2}{c}{M 5}
& \multicolumn{2}{c}{M53} & \multicolumn{2}{c}{M13}\\
\hline
  Y	 & \% &  Y	 & \% &  Y	 & \% &  Y	 & \%  &  Y	 & \%  &  Y	 & \%  &  Y	 & \% &  Y	 & \% \\
\noalign{\smallskip}
\hline  
\noalign{\smallskip}
0.24	& 50 &	0.25      & 38	&.25	& 39	&.25	&75  & .24 & .50 & .24 & .40 &.24 &1.& .24 & 0.0\\
0.26-0.29 &	35&	0.27-0.35	&48&	0.27-0.35&	41	&0.27-.32	&25  & .26 & 50 & .26 & 60 & $>$.24 & 0. & .28 & 1.0\\
$\sim$0.4 &	15 &	$>$0.35 &14 &	$>$0.35 &	20 && \\		
\noalign{\smallskip}
\hline
\end{tabular}
}
\end{center}
\label{table1}
\end{table}

\subsection{M3 and the problem of RR Lyr}

The case of M3 is entirely different: it has a well populated red HB, variable region and
blue side (R, V and B samples), with no blue tails, so it has always 
been taken as the prototype of HBs: its
color distribution can be reproduced by assuming an average mass loss along the RGB, with a 
standard deviation $\sigma \sim 0.025$\Msun. Unfortuntely, the RR Lyr period distribution
{\it is not} so easily explained, as it is terribly peaked! \cite{castellani2005} realized 
that the only way to reproduce this peak was to reduce the dispersion in mass loss rate: 
unfortunately, they also had to add a different average mass loss, with a different 
spread, to account for the blue side of the HB! Of course, if we assume that the blue side 
is populated by helium rich stars, the conundrum is solved, as shown
by \cite[Caloi \& D'Antona, 2007b]{caloi-dantona2007b}. In our simulations, we can explain both the period
distribution and the color distribution along the HB for the R, V and B regions. This analysis poses another
problem: we find that the dispersion in mass loss along the red giant branch must be at most
$\sigma \sim 0.003$\Msun\ to be consistent with the period distribution. We now pose the question
whether this small dispersion is peculiar to M3 or we have always been mislead by the HB
morphology, when assuming dispersion in mass loss of some hundreths of \Msun.

\subsection{Preliminary analysis of a few other clusters}
We list in Table 1 the results of the synthetic HB computations, 
including also the preliminary analysis for some other clusters.
M~5 is similar to M3, but has a larger fraction of higher helium stars (the peak of the
number of stars is in the blue HB). M~53 is a massive cluster, but its 
entirely blue HB could
be explained by a single (normal) stellar population, maybe with a small tail of stars having 
slightly larger mass loss. We list in the table also M~13, for which the analysis implying
that only the second generation has survived in the cluster is taken from the relative
location of the red giant bump, the turnoff and the HB (\cite[Caloi \& D'Antona, 2005]{caloi-dantona2005}).

\section{How did the GCs form?}

Any model for the GC formation must be able to deal with this variety of results: there
are clusters with no self--enrichment, and clusters which might have lost entirely their
first generation. It is true that the most massive clusters have extreme helium
enhancements, but also moderately massive clusters show considerable degrees of
helium variations, and a {\it small} cluster like NGC~6397, which is apparently 
monoparametric, might be composed entirely of second generation stars: in fact the
nitrogen abundance of all its stars is severely enhances, like in CNO processed 
material.

The typical situation is than 50\% normal--helium stars and 50\% enhanced--helium
stars. It is almost obvious that the ejecta of a unique first stellar generation with 
a normal initial mass function (IMF) can not produce enough mass to give origin to
such a large fraction of second generation stars (see, e.g., the case made by Bekki
and Norris (2004), for the blue main sequence of $\omega$Cen).

The only solution to this problem is that the starting initial mass from which the first 
generation is born was MUCH LARGER than today's first generation remnant mass (a factor 10
to 20 larger), so that the processed ejecta of the first generation provide enough mass 
to build up the second one. It is possible that a cooling flow collects the gas 
in the core, and that the second generaton stars are
preferentially born there (D'Ercole et al., in preparation). 
This formation in the central cluster region may also by helped
if the massive (or intermediate mass stars) 
progenitors were already segregated in the core when the second generation is born, as 
suggested by the observations of young clusters.
When the long term evolution leads to the loss of the external parts of clusters, 
these are populated preferentially by the first generation stars and the desired ratio
of first to second generation can be achieved.  

In some sense, the study of chemical anomalies leads us to the idea that {\it practically 
the GC forms in the second stellar generation}, and the remnant first generation 
is only the "core" fraction of the much less concentrated and wider system whose 
winds, collected in a cooling flow, gave birth to the "peculiar" stars!
Should we then expect different velocity dispersions in first and second generation?
Would the second generation –at least in some cases- remain more concentrated 
than the first one? of course complete dynamical simulations of the fluid gas formation
phase, coupled with N--body simulations to describe the interaction of the two stellar
generations are needed to answer this question.

\begin{acknowledgments}
We would like to thank P. Ventura, A. D'Ercole and E. Vesperini for their support and
collaboration through this adventure of building up a new view of globular clusters.
We also thank the organizers of the IAU Symposium 246 for their successful effort to
prepare a scientific and environmentally attractive meeting.
\end{acknowledgments}

\end{document}